\documentclass[epj]{svjour}
\usepackage{graphics}
\usepackage{graphicx}
\usepackage{amsfonts}
\usepackage{amsmath}
\usepackage{amsbsy}
\usepackage{longtable}
\usepackage{bm}
\usepackage{hyperref}
\usepackage{float}
\begin{document}
\title{Dissipative hydrodynamics for multi-component systems}
\author{Andrej El\inst{1}, Ioannis Bouras\inst{1}, Christian Wesp\inst{1}, Zhe Xu \inst{2} \and
Carsten Greiner\inst{1} 
}                     
%
%
\institute{Institut fuer Theoretische Physik, Goethe Universitaet, Max-von-Laue Str. 1, 60438
Frankfurt am Main, Germany \and Department of Physics, Tsinghua University, Beijing 100084, China}
\date{Received: date / Revised version: date}
%
\abstract{
Second-order dissipative hydrodynamic equations for each component of
a multi-component system are derived using the entropy principle. Comparison of the solutions with
kinetic transport results demonstrates validity of the obtained equations. We demonstrate how the
shear viscosity of the total system can be calculated in terms of the involved cross sections and
partial densities. Presence of the inter-species interactions leads to a characteristic
time-dependence of the shear viscosity of the mixture, which also means that the shear viscosity of
a mixture cannot be calculated using the Green-Kubo formalism the way it has been done recently.
This finding is of interest for understanding of the shear viscosity of a quark-gluon-plasme
extracted from comparisons of hydrodynamic simulations with experimental results from RHIC and LHC.
} 
\maketitle
\section{Introduction}
\label{intro}
The deconfined state of QCD matter produced at the early stage of 
ultrarelativistic heavy-ion collisions at RHIC and LHC is a multi-component
system with quark and gluon degrees of freedom.
Large values of elliptic flow coefficient $v_2$, observed at RHIC
\cite{Adler:2003kt,Voloshin:2008dg} and LHC \cite{Aamodt:2010pa},
indicate that the produced quark-gluon plasma (QGP) is a nearly perfect
fluid. This has motivated rapid developments on relativistic dissipative 
hydrodynamic formalisms \cite{IS,Muronga:2003ta,Romatschke:2009im,Denicol:2010xn,Monnai:2010qp}. 
The value of the shear
viscosity to the entropy density ratio $\eta/s$ for the QGP at RHIC and LHC was extracted from
comparisons of hydrodynamic \cite{hydros,Niemi:2012ry} as well as kinetic transport \cite{Xu:2007jv}
calculations
with experimental data. All these hydrodynamic formalisms
are based on the assumption that the quark-gluon mixture can be regarded as an effective
one-component system, where $\eta/s$ is an external parameter characterizing
the dissipation in the system. For a one-component system the $\eta/s$ ratio can be calculated using
the Green-Kubo formalism \cite{Wesp:2011yy,Fuini:2010xz} as well as other systematic
approaches \cite{Reining:2011xn,El:2008yy}. One may ask the question whether analogous calculation
of the shear viscosity of a mixture is possible, i.e. whether a mixture behavior is equivalent to a
one-component system or rather not. These questions are of major interest for
investigation of the QGP properties.

In this paper we demonstrate that a standard one-component hydrodynamic description with a single
shear viscosity coefficient calculated by  e.g. Green-Kubo formalism in general cannot be applied to
a multi\-component system. We will explain this statement by deriving second-order
dissipative hydrodynamic equations for a multi\- - component system from the entropy principle. Our
approach differs from the one reported in Ref. \cite{Monnai:2010qp}, since we introduce separate
evolution equations and transport coefficients for each component of the mixture. We then
show that by summing-up equations for all components one can obtain an equation for the system as a
whole, which has a relaxation-type form characteristic for all the present hydrodynamic formalisms,
but the effective shear viscosity for the mixture is now related to the partial shear pressures of
its components and thus has a non-trivial time dependence which is not supported by the Green-Kubo
or, for this matter, any other formalism, in which an equilibrium state of matter is assumed. We
will confirm our findings by calculating the shear viscosity of a mixture using the
Green-Kubo formula and comparing solutions of the hydrodynamic equations with the ones from kinetic
transport calculations.

\section{Dissipative hydrodynamic formalism for a multi\- -component system}

We consider a mixture of $N$ particle species, for which we define a common velocity field $u^\mu$.
Neglecting bulk pressure and heat flow we can construct the total entropy current as \cite{Kranys}
\begin{equation}
\label{smu1}
s^\mu=\sum_{i=1}^N s_i^\mu=s_{eq} u^\mu - \sum_{i=1}^N \frac{\beta_i}{2T_i}
\pi_{i,\alpha\beta} \pi_i^{\alpha\beta} u^\mu \,,
\end{equation}
where $s_{eq}$ is the total entropy density in local equilibrium and $u^\mu$ 
is the hydrodynamic velocity. $T_i$ and $e_i$ are the temperature and
local energy density of the particle species $i$. In analogy to the one-component case
\cite{Muronga:2003ta,El:2008yy} one obtains
$\beta_i=(9/4e_i)$. $\pi_i^{\mu\nu}=T_i^{\mu\nu}-T_{i,eq}^{\mu\nu}$ is the \textit{shear stress
tensor}, which is, as long as heat flow and bulk pressure are neglected, the difference between the
energy-momentum tensor $T_i^{\mu\nu}$ and the equilibrium one. Equation (\ref{smu1}) is the
generalization of the entropy current for a one-component system ($N=1$), discussed for instance in
Refs. \cite{IS,Muronga:2003ta,El:2008yy}.

The total entropy production is then
\begin{equation}
\partial_\mu s^\mu = \sum_i \pi_{i,\alpha\beta} 
\left[ \frac{\sigma^{\alpha\beta}}{T_i} - 
\pi_i^{\alpha\beta} \partial_\mu \left( \frac{\beta_i}{2T_i} u^\mu \right) -
\frac{\beta_i}{T_i} u^\mu \partial_\mu \pi_i^{\alpha\beta} \right] \,,
\label{dmusmu1}
\end{equation}
with the \textit{shear tensor}
\begin{equation}
\sigma^{\mu\nu}=\nabla^{\langle\mu}u^{\nu\rangle}=\left(\frac{1}{2}(\Delta_{\alpha}^{\mu}\Delta_{
\beta}^{\nu}+\Delta_{\alpha}^{\nu}\Delta_{\beta}^{\mu})-\frac{1}{3}\Delta_{\alpha\beta}\Delta^{
\mu\nu} \right)\nabla^{\alpha}u^{\alpha}
\nonumber
\end{equation}
and $\Delta_{\alpha\beta}=g_{\alpha\beta}-u_\alpha u_\beta$ with
the metric $g_{\alpha\beta}={\rm diag} (1,-1,-1,-1)$. We have used conservation of the partial
particle flows and total energy-momentum tensor, $\partial_\mu N_i^\mu=0$ and $\partial_\mu
T^{\mu\nu}=0$, to obtain Eq. (\ref{dmusmu1}). The obtained equation for the entropy balance is again
a generalization of the one-component result \cite{Muronga:2003ta}.

According to the second law of thermodynamics the entropy production
is non-negative. A simple way to fulfill this is to make the terms in the
square bracket in Eq. (\ref{dmusmu1}) to be proportional to 
$\pi_i^{\alpha\beta}$, $[\cdots ]=\pi_i^{\alpha\beta}/(2\eta_i T_i)$.
Then the entropy production (\ref{dmusmu1}) has the following algebraic 
structure:
\begin{equation}
\partial_\mu s^\mu \overset{!}{=} \sum_{i=1}^N
\frac{\pi_{i,\alpha\beta}\pi_i^{\alpha\beta}}{2\eta_i T_i} \ge 0 \,.
\label{dmusmu2}
\end{equation}
This leads to the dynamical equation for each $\pi_i^{\alpha\beta}$:
\begin{equation}
u^\mu \partial_\mu \pi_i^{\alpha\beta} = 
-\frac{\pi_i^{\alpha\beta}}{2\eta_i\beta_i}
-\pi_i^{\alpha\beta} \frac{T_i}{\beta_i}
\partial_\mu\left(\frac{\beta_i}{2T_i} u^\mu\right) 
+\frac{\sigma^{\alpha\beta}}{\beta_i} \,,
\label{pi1}
\end{equation}
which is analogous to the equation introduced by Israel and Stewart for a one-component system
($N=1$) \cite{IS,Muronga:2003ta}.

In order to apply Eq. (\ref{pi1}) to a multi-component system ($N>1$), we first need to determine
the coefficients $\eta_i$, which in general differ from the usual definition of the shear viscosity.
The reason for this is that $\pi_i^{\alpha\beta}$'s are \textit{correlated} due to
interactions between particles from different species.
These correlations between $\pi_i^{\alpha\beta}$'s can only be seen, when
each $\eta_i$ depends on all $\pi_j^{\alpha\beta}$, $j=1,2, \cdots, N$.
We will also show later that the coefficients $\eta_i$ become the shear viscosities, when
the ratios of components of $\pi_i^{\alpha\beta}$'s are relaxing to constants in time.

We now make use of relativistic kinetic theory and express the entropy current 
via the phase-space distribution function $f_i(x,p_i)$:
\begin{equation}
\label{smu2}
s^\mu=\sum_{i=1}^N \int d\Gamma_i \ p_i^\mu f_i(x,p_i)\, [1-\ln f_i(x,p_i)]
\end{equation}
with $d\Gamma_i=d^3p_i/E_i/(2\pi)^3$. It was shown for the case of 
$N=1$ \cite{3rdO} and is obviously true for $N>1$ that using the Grad's 
ansatz \cite{Muronga:2003ta}
\begin{equation}
f_i(x,p_i) = f_{i,eq}(x,p_i) ( 1 + A_i \pi_{i,\mu\nu} p_i^\mu p_i^\nu) 
\label{eq:grad}
\end{equation}
in Eq. (\ref{smu2}) one obtains Eq. (\ref{smu1}) up to second order
in $\pi_{i,\mu\nu}$. Here $f_{i,eq}(x,p_i)$ is the equilibrium distribution
function and $A_i= [ 2 (e_i + P_i) T_i^2 ]^{-1}$ \cite{El:2008yy}, where $P_i$ is the pressure.

The space-time evolution of $f_i(x,p_i)$ obeys the Boltzmann equation
\begin{equation}
\label{boltzmann}
p_i^\mu \partial_\mu f_i = C_i[f_1, f_2, \cdots, f_N]= 
C_{ii}[f_i] +  \sum_{j=1,j\ne i}^N C_{ij}[f_i,f_j] \,,
\end{equation}
where $C_{ii}$ are the collision terms describing interactions of
particles of same species and $C_{ij}$ describing binary interactions of
particles of different species. Explicit expressions for the collision 
terms can be found for example in \cite{Xu:2004mz}. Taking derivative 
of (\ref{smu2}) and using (\ref{boltzmann}) we obtain
\begin{equation}
\label{dmusmu3}
\partial_\mu s^\mu = \sum_{i=1}^N A_i \pi_{i,\mu\nu} 
\int d\Gamma_i \, p_i^\mu p_i^\nu C_i \,.
\end{equation}
Comparison between Eqs. (\ref{dmusmu3}) and (\ref{dmusmu2}) leads to
\begin{equation}
\eta_{i} = \frac{\pi_{i,\mu\nu} \pi_i^{\mu\nu}}
{2A_i \pi_{i,\mu\nu} \int d\Gamma_i p_i^\mu
p_i^\nu  C_i} \,.
\label{eta_gen}
\end{equation}
Because the collision term $C_i$ is a functional of all $f_j$'s,
each $\eta_i$ depends on all $\pi_j^{\mu\nu}$'s with $j=1,2,\cdots,N$.

In order to simplify Eq. (\ref{eta_gen}) we will now consider a one-dimensional system, which
implies that the shear stress tensor has a diagonal (and of course traceless) form:
$\pi^{\alpha\beta} =
diag(0,\pi/2,\pi/2,-\pi)$, with the single independent component $\pi$. By virtue of Eq.
(\ref{eq:grad}) this form of the shear stress tensor is equivalent to a deformation of the
momentum-space distribution along the $z$ axis, whereas in the transverse $xy$ plain the momentum
distribution is isotropic. Furthermore we consider only two species, $i,j={1,2}$ and isotropic
scattering processes, i.e. we assume that the differential cross section $d\sigma/d\Omega$ does not
depend on scattering angle. Inserting the
off-equilibrium destribution (\ref{eq:grad}) into the collision term in (\ref{eta_gen}) and using
the aforementioned simplifications we obtain the following expression for the shear viscosities of
the mixture constituents:
\begin{equation}
\eta_{i}^{-1}  = T_i^{-1} \sum_{j=1}^N \left (\frac{7}{6}\frac{n_j}{n_i}
- \frac{1}{3}\frac{\pi_j}{\pi_i} \right ) \sigma_{ij} \,.
\label{eta_iso} 
\end{equation}
In the latter equation $n_j/n_i$ denotes the ratio of the particle densities of mixture
constituents. In our formalism this chemical composition is fixed by the initial consition and does
not change in time, although this assumption is a very strong simplification for a QGP. Now
the obtained expression for $\eta_i$ can be inserted into the dynamic evolution equation (\ref{pi1})
for $\pi_i$ (that is we take the $\pi_{33}$ component of $\pi_{\mu\nu}$ in (\ref{pi1}), although
considering any other component will lead to identical result). We obtain a dynamic equation for
the shear tensor components $\pi_i$ in a mixture:
\begin{align}
\dot \pi_i = &- \left( \frac{5}{9} n_i\sigma_{ii} + \frac{7}{9} n_j\sigma_{ij} \right) \pi_i +
\frac{2}{9} n_i \sigma_{ij} \pi_j \nonumber \\
&-\pi_i\frac{T_i}{\beta_i}\partial_\mu \left( \frac{\beta_i}{2T_i} u^\mu \right) +
\frac{\sigma}{\beta_i} \,\label{pi_i}
\end{align}
with the short-hand notation $\dot \pi \equiv u_\mu \partial^\mu \pi$ and $\sigma$ denoting the $zz$
component of the shear tensor $\sigma=\sigma^{33}$. A distinct feature of the obtained
equation, which has a characteristic relaxation-time form, is the presence of two time scales for
relaxation, of which the second one is associated with a coupling between the partial shear
pressures $\pi_i$ of the two species. 

\section{Coupled dynamics in a mixture}

Now the question arises, as of how the dependence of $\eta_i$ in Eq. (\ref{eta_iso}) on the ratio
$\pi_i/\pi_j$ can be interpreted. If $\eta_i$ is understood as the shear viscosity of the medium,
it must be a property of the medium, i.e. depend exclusively on its chemical composition and the
associated cross sections. Moreover, the shear viscosity must be defined in the proximity of the
equilibrium. Let us consider the following situation: all velocity gradients vanish. This means
that the shear tensor $\sigma^{\mu\nu}$, and thus $\sigma$ as well, vanish. We also assume the the
temperatures of the two sub-systems are equal and can be replaced by the single temperature $T$ of
the system, i.e. $T_i=T$. We move into the fluid
rest frame, in which $u_\mu \partial^\mu \equiv \partial/\partial\tau$. Equations
(\ref{pi_i}) are now reduced to a set of two coupled differential equations for the shear stress
tensor
components $\pi_1$, $\pi_2$ with the parameters $n_1$, $n_2$, $\sigma_{11}$, $\sigma_{12}$ and
$\sigma_{22}$. The
system of equations can be solved analytically and thus a solution for $\frac{\pi_1}{\pi_2}(\tau)$
is also found:
\begin{equation}
\frac{\pi_1}{\pi_2}(\tau) = \mathcal{A}(n,\sigma)\cdot \tanh \left( \tau\cdot\mathcal{B}(n,\sigma) 
+
\mathcal{D}(n,\sigma,\pi_0)\right) \,. 
\label{eq:pi1pi2}
\end{equation}
In the latter equation $\mathcal{A}$,  $\mathcal{B}$ and $\mathcal{D}$ are algebraic functions of
the system properties $n_i$ and $\sigma_{ij}$, and in particular $\mathcal{D}$ depends on the
initial value of the ratio $\pi_1/\pi_2$. Equation (\ref{eq:pi1pi2}) demonstrates that the ratio
$\pi_1/\pi_2$ of the partial shear pressures is determined completely by the properties of the
system, but is not a constant. Moreover, Eq. (\ref{eq:pi1pi2}) leads to saturation of $\pi_1/\pi_2$
in $\tau\to\infty$ limit, i.e. a well-defined characteristic limit for the ratio exists. In general
this characteristic value is different from the value $n_1/n_2$. It is also important to mention,
that there is no conclusive way to specify the initial values of $\pi_1$ and $\pi_2$ -- or just
$\pi$ for a standard one-component hydrodynamic calculation. In most hydrodynamic
approaches the standard choice is therefore the trivial initialization $\pi(\tau_0) = 0$. Another
possible choice is initialization with the Navier-Stokes value $\pi(\tau_0)=2\eta\sigma$
\cite{Niemi:2012ry}. As it is for the one-component case, for a multi-component system the choice
of the initial condition for $\pi_i$ is not clear as well. If the gradients are 'switched on', the
trivial choice $\pi_i(\tau_0)=0$ will lead, according to Eq. (\ref{pi_i})
to $\dot\pi_i=\sigma/\beta_i$. And since $\beta_2/\beta_1=e_1/e_2=n_1/n_2$, after a short time the
shear pressure ratio will be approximately equal to the density ratio, $\pi_1/\pi_2\approx n_1/n_2$.

It is also interesting to build the sum of Eqs. (\ref{pi_i}), which leads us to a relaxation
equation for the total shear pressure in the mixture $\dot \pi = \dot\pi_1+\dot\pi_2$. If we write
this equation in the relaxation-time form with only one relaxation scale, as is usual for most
hydrodynamic formalisms presently used \cite{hydros,Niemi:2012ry}, we obtain
\begin{align}
\dot \pi = &-\pi \cdot \frac{5}{9}\bigg(
\frac{\pi_1/\pi_2}{1+\pi_1/\pi_2}\cdot(n_1\sigma_{11} +n_2\sigma_{12}) \nonumber\\
&+\frac{1}{1+\pi_1/\pi_2}\cdot(n_2\sigma_{22} + n_1\sigma_{12}) \bigg) + gradients
\,.\label{totalpi}
\end{align}
According to the standard definition of the relaxation time $\tau_\pi$ in the second-oder
formalisms \cite{IS,Muronga:2003ta}
\begin{equation}
 \tau_\pi = \frac{9\eta}{2e}
\end{equation}
we recognize that the shear viscosity of the mixture is now given by
\begin{equation}
\eta_{mix} = \frac{2}{5} e \bigg( \frac{\pi_1/\pi_2}{1+\pi_1/\pi_2 }\cdot\lambda_1^{-1} +
\frac{1}{1+\pi_1/\pi_2}\cdot \lambda_2^{-1} \bigg)^{-1} \,,  
\label{eta_mix}
\end{equation}
with the inverse of the mean free path $\lambda_1^{-1} = n_1\sigma_{11} + n_2\sigma_{12}$ and
$\lambda_2^{-1} = n_2\sigma_{22} + n_1\sigma_{12}$ and the time-dependent ratio $\pi_1/\pi_2$ given
by Eq. (\ref{eq:pi1pi2}). 

The obtained result for the shear viscosity of the mixture is interesting for the following reason.
If one attempts to calculate the shear viscosity of the mixture we have considered here, e.g.
using the Green-Kubo formula in a kinetic transport simulation, the result would naturally be a
constant value for $\eta_{mix}$. On the other hand the result obtained here in form of Eq.
(\ref{eta_mix}) implies that the viscosity of a mixture \textit{is} time-dependent, but saturates.
The value at which the time-dependence dies off will \textit{not} be identical with the value
one would obtain using the Green-Kubo formalism because the coupling between the species in a
mixture induces an internal dynamics in a system.

\begin{figure}
\resizebox{0.5\textwidth}{!}{%
  \includegraphics{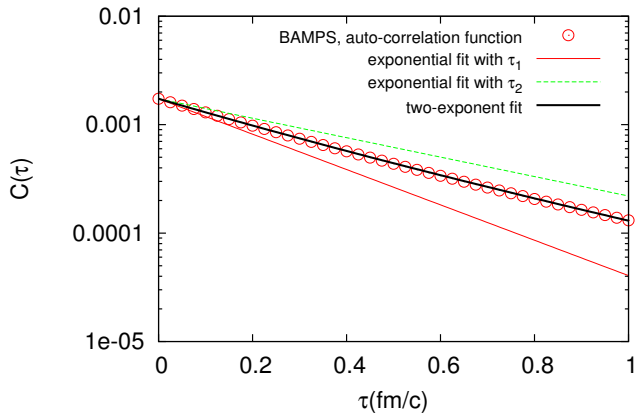}
}
\caption{Autocorrelation function extracted from BAMPS (symbols) as function of correlation time.
Two-exponent fit Eq. (\ref{fit}) with $C_1=C_2=8.66\cdot 10^{-4} GeV^2/fm^6$,
$\tau_1=0.264822~fm/c$ and $\tau_2=0.479556~fm/c$ is shown by bold solid line. The two
exponents from the two-exponent fit are also shown separately by the thin solid and dashed lines.}
\label{fig:acf}      
\end{figure}

\begin{figure}
\resizebox{0.5\textwidth}{!}{%
  \includegraphics{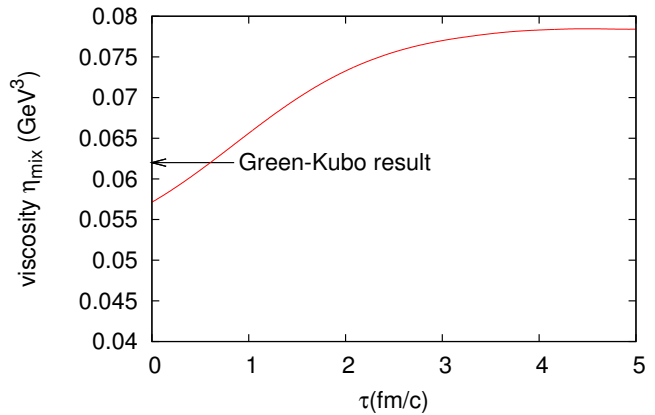}
}
\caption{Time evolution of the shear viscosity $\eta_{mix}$ from Eq. (\ref{eta_mix}). See the text
for details of the setup. Result of the Green-Kubo formalism is indicated by the arrow.}
\label{fig:eta}      
\end{figure}

To demonstrate this, we calculate the shear viscosity coefficient of a mixture in the kinetic
transport model Boltzmann Approach to MultiParton Scatterings (BAMPS) \cite{Xu:2004mz,Xu:2007jv}
using the procedure successfully applied by us in Ref. \cite{Wesp:2011yy}. The cross sections for
the scattering processes of the two species, confined in a static box, are chosen to be
$\sigma_{11}=10~GeV^{-2}$, $\sigma_{12}=5~GeV^{-2}$ and $\sigma_{22}=2.5~GeV^{-2}$. The density
ratio is $n_1/n_2=5$. The temperature is chosen to be $T=0.4~GeV$ and both particle species are
considered to be Boltzmann gases with degeneracy factors $16$, i.e. $n_1=5/6\cdot 16/\pi^2 T^3$ and
$n_2=1/6\cdot 16/\pi^2 T^3$. Equation of state is the ideal one, i.e. $e_1=3 n_1 T$ and $e_2=3 n_2
T$. Note that for this setup the mean-free path
scales for the two species are $\lambda_1 = 0.207~fm$ and $\lambda_2 = 0.414~fm$. These values are
chosen to crudely simulate quarks and gluons in a QGP. To obtain the shear viscosity, we extract the
correlation function 
\begin{equation}
C(\tau)=\frac{1}{3} \left( \langle \pi^{xy}(0) \pi^{xy}(\tau)\rangle + \langle \pi^{xz}(0)
\pi^{xz}(\tau)\rangle +  \langle \pi^{yz}(0)
\pi^{yz}(\tau)\rangle \right)
\end{equation}
where $\tau$ is the correlation time and $\langle . \rangle$ denotes ensemble-average in the static
box. The Green-Kubo formalism relates the shear viscosity to the integral of the correlation
function over the relaxation time:
\begin{equation}
 \eta = \frac{V}{T} \int\limits_{0}^\infty C(\tau) d\tau \label{etaGK} \,,
\end{equation}
where $V$ denotes the considered volume and $T$ the temperature of the system.
In a standard one-component case the correlator $C(\tau)$ is very good described by an
exponential function with the relaxation time $\tau_c$ \cite{Wesp:2011yy,Fuini:2010xz}. For a
mixture, however, we find that a single-exponent fit does not work and two-exponent fit must be
considered:
\begin{equation}
C(\tau) = C_1\cdot e^{-\tau/\tau_1} + C_2\cdot e^{-\tau/\tau_2} \,,\label{fit}
\end{equation}
with two relaxation times $\tau_1$ and $\tau_2$. In Fig. \ref{fig:acf} we demontrate the
correlation function extracted from BAMPS static box calculations together with the two-exponetial
fit. Integrating the correlation function shown in Fig. \ref{fig:acf} over the correlation time we
obtain for the shear viscosity $\eta=0.062~GeV^{-3}$. To demonstrate the difference between this
result and the effective mixture viscosity (\ref{eta_mix}) we show in Fig. \ref{fig:eta} the time
evolution of $\eta_{mix}$ for the setup described above. From Fig. \ref{fig:eta} we recognize that
the mixture viscosity $\eta_{mix}$ is approximately equal to the Green-Kubo result at erly evolution
stage but significantly increases with time.
\begin{figure}
\resizebox{0.5\textwidth}{!}{%
  \includegraphics{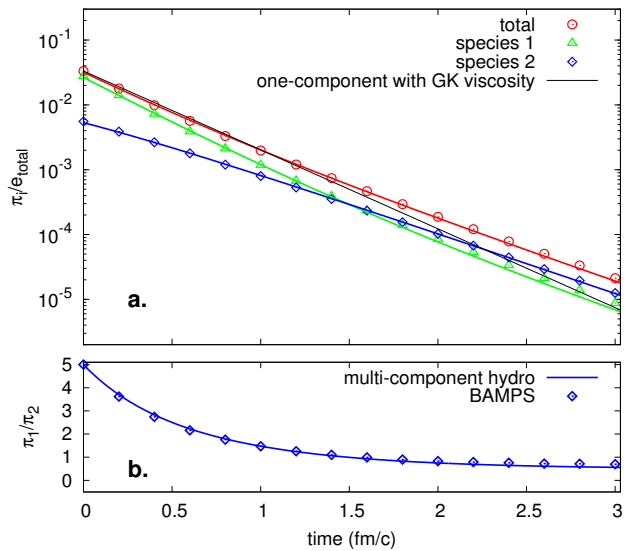}
}
\caption{Relaxation of the total and partial shear pressures in BAMPS (symbols) and analytic
solutions (lines). Additionally, one-component solution with the viscosity obtained using the
Green-Kubo formalism is shown to reproduce the  total shear pressure of a multi-component system
only at early times of the evolution.}
\label{fig:relax}      
\end{figure}
In order to verify the applicability of the obtained relaxation-type equations (\ref{pi_i}) for the
shear pressure we now study relaxation of shear pressure in a static BAMPS box. We use
a quasi-static setup, i.e. a volume with no gradients but finite initial shear pressures $\pi_1(0)$
and $\pi_2(0)$. In the kinetic
transport solver BAMPS this is achieved by sampling particles isotropically in space according to
the distribution function (\ref{eq:grad}) with a chosen value of $\pi=\pi^{zz}$. This setup provides
a cross-check of the relaxation dynamics described by the first two terms on the right-hand-side of
Eqs. (\ref{pi_i}) and also provides a possibility to cross-check of the validity of Eq.
(\ref{eq:pi1pi2}). We use same cross sections and composition of the gas as for the Green-Kubo
calculations discussed above. Results of BAMPS static box calculations are shown in
Fig. \ref{fig:relax}. The BAMPS results (symbols) for the total as well as the partial shear
pressures (Fig. \ref{fig:relax} (a)) are well reproduced by the analytic solutions of Eq.
(\ref{pi_i}). The ratio $\pi_1/\pi_2$ calculated in BAMPS is also shown to agree with the analytic
solution and demonstrates the expected saturation (Fig. \ref{fig:relax} (b)). The one-component
solution $\pi(\tau) = \pi (\tau_0)\cdot e^{-\tau/\tau_\pi}$ with the shear viscosity value
$\eta_{GK}$ calculated using Green-Kubo formalism is shown by the solid black line in Fig.
\ref{fig:relax} (a) and is unable to describe relaxation of the total shear pressure in a
multi-component system on a long time scale.  This means that the Green-Kubo formalism in the form
it is applied to one-component systems cannot be applied to calculate the shear viscosity of a
mixture. From Fig. \ref{fig:relax} (a) we rather recognize that the viscosity must increase with
time, though it might be close to the Green-Kubo result at early times of the evolution.

\section{Conclusions}

We have derived second-order hydrodynamic equations for the shear tensor components of
constituents of a mixture. A cross-check of the obtained equations is provided by comparisons of
the solutions with kinetic transport calculations with BAMPS, which demonstrate very good agreement
of the results. We have also demonstrated that the effective shear viscosity of a mixture of two
components does have a non-trivial time-dependence, which is explained by inner dynamics of the
mixture due to inter-species interactions. If the Green-Kubo formalism is applied to calculate the
shear viscosity of a mixture, the result cannot capture its time-dependence. Thus, if calculated by
the Green-Kubo formalism, the shear viscsosity cannot be used to describe hydrodynamic evolution of
a mixture. It will be very interesting to investigate the impact of our finding on extraction of the
shear viscosity of a Quark-Gluon mixture from comparisons of experimental data with the
results of dissipative hydrodynamic simulations. We expect for example, that the elliptic flow
coefficient $v_2(p_T)$ calculated from a mixture, such as the Quark-Gluon plasma, cannot be
accurately reproduced by a one-component dissipative hydrodynamic calculations unless the proper
time-dependence of the shear viscosity, which we have introduced in this paper, is taken into
account.

\section*{Acknowledgements}
This work was supported by the Helmholtz International Center for FAIR
within the framework of the LOEWE program launched by the State of Hesse. IB and CW acknowledge
support by HGS-Hire. The authors are greatful to the Center for Scientific Computing (CSC) at
Frankfurt University and LOEWE-CSC for the computing resources


\begin{thebibliography}{}

\bibitem{Adler:2003kt}
  S.~S.~Adler {\it et al.} [ PHENIX Collaboration ],
  Phys.\ Rev.\ Lett.\  {\bf 91}, 182301 (2003).
  [nucl-ex/0305013].

\bibitem{Voloshin:2008dg}
  S.~A.~Voloshin, A.~M.~Poskanzer and R.~Snellings,
  arXiv:0809.2949 [nucl-ex].

\bibitem{Aamodt:2010pa}
  K.~Aamodt {\it et al.}  [The ALICE Collaboration],
  arXiv:1011.3914 [nucl-ex].

\bibitem{IS}
W.~Israel, Ann.\ Phys.\ (N.Y.)  100 (1976) 310;
J.M.~Stewart, Proc.\ R.\ Soc.\ London\ , Ser.\ A\  357 (1977) 59;
W.~Israel, M.~Stewart, Ann.\ Phys.\ (N.Y.) 118 (1979) 341.

\bibitem{Muronga:2003ta}
  A.~Muronga,
  Phys.\ Rev.\  {\bf C69}, 034903 (2004).
  [arXiv:0309055 [nucl-th]].

\bibitem{Romatschke:2009im}
  P.~Romatschke,
  Int.\ J.\ Mod.\ Phys.\  {\bf E19}, 1-53 (2010).
  [arXiv:0902.3663 [hep-ph]].

\bibitem{Denicol:2010xn}
  G.~S.~Denicol, T.~Koide, D.~H.~Rischke,
  Phys.\ Rev.\ Lett.\  {\bf 105}, 162501 (2010).
  [arXiv:1004.5013 [nucl-th]].

\bibitem{Monnai:2010qp}
  A.~Monnai, T.~Hirano,
  Nucl.\ Phys.\  {\bf A847}, 283-314 (2010).
  [arXiv:1003.3087 [nucl-th]].

\bibitem{hydros}
  H.~Song and U.~W.~Heinz,
  Phys.\ Rev.\   {\bf C77}, 064901 (2008)
  [arXiv:0712.3715 [nucl-th]];

  M.~Luzum and P.~Romatschke,
  Phys.\ Rev.\  C {\bf 78}, 034915 (2008)
  [Erratum-ibid.\  C {\bf 79}, 039903 (2009)]
  [arXiv:0804.4015 [nucl-th]];

  D.~A.~Teaney,
  [arXiv:0905.2433 [nucl-th]];

  H.~Song, S.~A.~Bass and U.~W.~Heinz,
  arXiv:1103.2380 [nucl-th].

\bibitem{Niemi:2012ry} 
  H.~Niemi, G.~S.~Denicol, P.~Huovinen, E.~Molnar and D.~H.~Rischke,
  arXiv:1203.2452 [nucl-th].


\bibitem{Xu:2007jv}
  Z.~Xu, C.~Greiner and H.~St\"ocker,
  Phys.\ Rev.\ Lett.\  {\bf 101}, 082302 (2008)
  [arXiv:0711.0961 [nucl-th]].

\bibitem{Wesp:2011yy} 
  C.~Wesp, A.~El, F.~Reining, Z.~Xu, I.~Bouras and C.~Greiner,
  Phys.\ Rev.\ C {\bf 84}, 054911 (2011)
  [arXiv:1106.4306 [hep-ph]].

\bibitem{Fuini:2010xz}

  N.~Demir and S.~A.~Bass,
  Phys.\ Rev.\ Lett.\  {\bf 102}, 172302 (2009)
  [arXiv:0812.2422 [nucl-th]];

  J.~I.~Fuini, N.~S.~Demir, D.~K.~Srivastava and S.~A.~Bass,
  J.\ Phys.\ G {\bf 38}, 015004 (2011)
  [arXiv:1008.2306 [nucl-th]].

\bibitem{Reining:2011xn} 
  F.~Reining, I.~Bouras, A.~El, C.~Wesp, Z.~Xu and C.~Greiner,
  Phys.\ Rev.\ E {\bf 85}, 026302 (2012)
  [arXiv:1106.4210 [hep-th]].

\bibitem{El:2008yy} 
  A.~El, A.~Muronga, Z.~Xu and C.~Greiner,
  Phys.\ Rev.\ C {\bf 79}, 044914 (2009)
  [arXiv:0812.2762 [hep-ph]].


\bibitem{Kranys} Krany{\v s}, M.\ 1970, 
Archive for Rational Mechanics and Analysis, 39, 245 .

\bibitem{3rdO}
  A.~El, Z.~Xu and C.~Greiner,
  Phys.\ Rev.\  C {\bf 81}, 041901 (2010)
  [arXiv:0907.4500 [hep-ph]].

\bibitem{Xu:2004mz}
  Z.~Xu, C.~Greiner,
  Phys.\ Rev.\  {\bf C71}, 064901 (2005).
  [arXiv:0406278 [hep-ph]];

  Phys.\ Rev.\  C {\bf 76}, 024911 (2007)
  [arXiv:hep-ph/0703233].

\bibitem{HM09}
  P.~Huovinen and D.~Molnar,
  Phys.\ Rev.\  C {\bf 79}, 014906 (2009)
  [arXiv:0808.0953 [nucl-th]].

\bibitem{Denicol:2011fa}
  G.~S.~Denicol, J.~Noronha, H.~Niemi and D.~H.~Rischke,
  [arXiv:1102.4780 [hep-th]].

\bibitem{DeGroot}
S.~R. ~de Groot, W. ~A. ~van Leeuwen, Ch.~G.~van Weert, Relativistic Kinetic Theory: Principles and
Applications, North Holland, Amsterdam, 1980.

\bibitem{Bouras:2010hm}
  I.~Bouras {\it et al.},
  Phys.\ Rev.\ Lett.\  {\bf 103}, 032301 (2009)
  [arXiv:0902.1927 [hep-ph]];

  Phys.\ Rev.\  {\bf C82}, 024910 (2010).
  [arXiv:1006.0387 [hep-ph]].























\end{thebibliography}
\end{document}